\documentclass{iau} 
\usepackage{graphicx}

\title[Fundamental Physics and Cosmology in the ELTs Era] %% give here short title %%
{Fundamental Physics and Cosmology in the Extremely Large Telescopes Era}
\author[C. J. A. P. Martins]   %% give here short author list %%
{C. J. A. P. Martins}
\affiliation{Centro de Astrof\'{\i}sica, Universidade do Porto, \\[\affilskip]
Rua das Estrelas, 4150-762 Porto, Portugal \\[\affilskip]
and Instituto de Astrof\'{\i}sica e Ci\^encias do Espa\c co, Universidade do Porto, \\[\affilskip]
Rua das Estrelas, 4150-762 Porto, Portugal \\ email: {\tt Carlos.Martins@astro.up.pt}}

\pubyear{2018}
\volume{347}  %% insert here IAU Symposium No.
\jname{Early Science with ELTs}
\editors{A.C. Editor, B.D. Editor \& C.E. Editor, eds.}
\begin{document}

\maketitle

\begin{abstract}
The observational evidence for the recent acceleration of the universe demonstrates that canonical theories of cosmology and particle physics are incomplete (or possibly incorrect) and that new physics is out there, waiting to be discovered. A crucial task for the next generation of astrophysical facilities is to search for, identify and ultimately characterise this new physics. I outline the theoretical arguments pointing to this new physics, and then discuss some unique contributions of the ELTs towards this endeavour, including tests of the stability of nature’s fundamental couplings, tests of the behaviour of gravity in the strong-field regime, and mapping the expansion history of the universe in the deep matter era, by both direct and indirect means. I also briefly highlight how the ELTs can optimally complement other planned ground and space facilities, and stress the requirements that these searches impose on the instruments currently being built or developed.
\end{abstract}

\firstsection % if your document starts with a section,
\section{Introduction}

Fundamental physics and cosmology are at the dawn of a revolution. The evidence for the acceleration of the universe shows that their canonical theories are at least incomplete, and possibly incorrect. Observationally we know that the component responsible for this acceleration has a gravitational behaviour very similar to that of a cosmological constant (or equivalently, vacuum energy). Whether or not this is the case, the implications are dramatic. If the answer is yes, this vacuum energy is many orders of magnitude smaller than expected in Quantum Field Theories. If the answer is no, any plausible alternative mechanism will violate the Einstein Equivalence Principle.

One way or the other, at the most fundamental level our current view of the Universe is incomplete. New physics is out there, waiting to be discovered. Thus the most compelling task of the next generation of experimental and observational facilities is to search for, identify and ultimately characterise this new physics. In this contribution I will highlight the unique role of the ELTs in this quest. Given the limited time of the invited talk (and the limited space of the present written version) I will mostly focus on the science cases relying on high-resolution spectroscopy (thus in the scope of the ELT-HIRES instrument), but I will also briefly mention other instruments, as well as other facilities. Full disclosure: I am a member of ESO's ELT Project Science Team, the ESPRESSO and ELT-HIRES Science Teams, and also Euclid and LISA.

It is worth starting by defining the term {\it fundamental physics}. For our present purposes, this comprises two distinct but nevertheless inter-related aspects: tests of fundamental laws and symmetries (e.g., tests of the Equivalence Principle in its various forms, or of the behaviour of gravity on all scales), and searches for Nature’s fundamental constituents (including scalar fields as an explanation for dark energy, new particles for dark matter, magnetic monopoles or fundamental strings. Importantly, many of these principles are necessarily violated in extensions of the standard model: the spacetime structure is modified (violating Lorentz Invariance), fundamental couplings become dynamical, violating the Einstein Equivalence Principle (which we will discuss in detail in what follows), and gravity laws are modified at large and/or small scales.

Fundamental scalar fields are particularly relevant, because we know (thanks to the LHC) that they are among Nature's building blocks. These fields will naturally couple to the rest of the model, unless there is an unknown principle to suppress them. These couplings will therefore lead to potentially observable long-range forces and varying couplings, as discussed in \cite[Carroll (1998)]{Carroll}. Improved measurements of these couplings (whether they are detections or null results) constrain fundamental physics and cosmology. This ensures a 'minimum guaranteed science' for the forthcoming observational facilities. These issues are further described in \cite[Martins (2017)]{ROPP}.

\section{Varying fundamental couplings}

Nature is characterised by some physical laws and dimensionless couplings, which historically  we have assumed to be spacetime-invariant. For the former, this is a cornerstone of the scientific method (it's hard to imagine how one could do science otherwise), but for latter it is merely a simplifying assumption without further justification. We have no {\it theory of constants} describing their role in physical theories---and if they vary, all the physics we know is incomplete. Improved null results are important and theoretically very useful, while a detection of variations would be revolutionary: among other consequences, varying dimensionless physical constants imply a violation of the Einstein Equivalence Principle and a fifth force of Nature. Currently the observational status is somewhat unclear, and limited by the lack of sufficiently precise instruments such as high-resolution ultra-stable spectrographs---see \cite[Martins (2017)]{ROPP} for a recent review. However, forthcoming facilities such as the ELTs will lead to dramatic progress in the field.

It is worth emphasising a key point: if no variations are seen at a certain level of sensitivity, should one make an effort to tighten these bounds? An analogy with dynamical dark energy provides the clearest way to understand the answer. Let’s consider the present-day value of the dark energy equation of state, $w_0$, or more specifically $(1 + w_0)$ which is the dynamically relevant quantity (for a canonical scalar field this is just the ratio of the field's kinetic and total energies. Naively we would expect this to be of order unity, but observationally we know that it must (conservatively) be less than 0.1. The point is that if this number is not of order unity there is no natural scale for it: either there is some fine-tuning to make it small, or there is a new (currently unknown) symmetry which forces it to be zero. An analogous argument holds for relative variations of dimensionless couplings, such as the fine-structure constant or the proton-to-electron mass ratio, the only difference being that the must be less than must be less than $10^{-5}$.

Thus if no variations are confirmed at the $10^{-6}$ level (the sensitivity of spectrographs such as ESPRESSO), should the ELTs push even further? Certainly the answer is yes, and the Strong CP Problem in QCD clearly illustrates why: a parameter which we would have expected (given current knowledge of particle physics) to be of order unity is actually smaller than $10^{-10}$, leading to the postulate of the Peccei-Quinn symmetry and a range of further interesting consequences. A sufficiently tight bound will either imply that there are no dynamical scalar fields fields in cosmology or that its couplings to the rest of the model are suppressed by some currently unknown symmetry of Nature---whose existence would be at least as significant as that of the original field.

In theories where a dynamical scalar field leads to a varying fine-structure constant, not only does one expect other couplings such as the proton-to-electron mass ration to vary, but there must also be impacts in other cornerstones of the standard cosmological model. Notably, there will be violations of the temperature-redshift law (which canonically assumes adiabatic expansion and photon number conservation) and of the distance duality relation (also known as Etherington relation, which canonically assumes a metric theory of gravity and photon number conservation). These relations therefore provide a unique opportunity for fundamental consistency tests of the standard cosmological model: current joint constraints are at the $0.8\%$ level---see \cite[Avgoustidis \etal\ (2016)]{Avgoustidis}---and are expected to improve in the future. For this reason, astrophysical targets where several parameters can be measured simultaneously are particularly relevant from a theoretical point of view.

As for measurements of the proton-to-electron mass ratio, the most common way of astrophysically measuring is using vibrational and rotational transitions of molecular hydrogen. Analogous measurements in other molecules (which are often far more sensitive to possible variations than $H_2$ itself) one is actually measuring a ratio of an effective nucleon mass to the electron mass, and its relative variation of this quantity will only equal that of the proton-to-electron mass ratio if there are no composition-dependent forces: in other words, if protons and neutrons have identical couplings to the putative scalar fields. This provides a golden opportunity to the ELTs: stringent astrophysical constraints on composition-dependent forces can by carried out by finding a system where relevant vibrational and rotational transitions can be separately measured from different molecules with different numbers of protons and neutrons: for example $H_2$, $HD$, and perhaps also carbon monoxide, ammonia or methanol which are all (comparatively) common molecules. This will be a revolutionary direct astrophysical test of the Weak Equivalence Principle. Apart from identifying a suitable astrophysical system (searches for such targets are ongoing) the main bottleneck is a suitable wavelength coverage in the blue part of the spectrum---an issue to which we will return.

\section{Implications for dark energy}

The Universe is seemingly dominated by a component whose gravitational behaviour is similar to that of a cosmological constant. A cosmological constant may indeed be behind it, but given the well-known problems associated with such a solution, a dynamical scalar field is (arguably) more likely. Such a field must be presently slow-rolling (which is mandatory for $p<0$) and be dominating the dynamics around the present day. These are sufficient to ensure that couplings of this field will lead to potentially observable long-range forces and varying fundamental couplings, as discussed in \cite[Carroll (1998)]{Carroll}. In what follows we will illustrate how current measurements already provide competitive constraints on fundamental physics and cosmology. This explains why these tests are flagship science cases (and design drivers) for forthcoming astrophysical facilities including the ELTs.

Any scalar field couples to gravity; it couples to nothing else if a global symmetry suppresses couplings to the rest of the Lagrangian (in which case only derivatives and derivative couplings survive). However quantum gravity effects do not respect global symmetries and there are no unbroken global symmetries in string theory. From a fundamental physics perspective, the most natural scenario is therefore that the same scalar degree of freedom yields dynamical dark energy and a varying fine-structure constant. It follows that in the simplest models (including, but not limited to, quintessence-type models) the latter's evolution is parametrically determined. A combination of astrophysical and local measurements of the fine-structure constant and background cosmology data (such as Type Ia supernova and Hubble parameter data) then constrains the coupling $\zeta$ of the scalar field to the electromagnetic sector of the theory. The current best constrains are $|\zeta|<4\times10^{-6}$---see \cite[Martins \etal\ (2016)]{Pinho}---at the $95.4\%$ confidence level, which is further discussed in \cite[Martins (2017)]{ROPP}. With the recently started ESPRESSO GTO this bound is expected to improve by an order of magnitude (for null results), or a non-zero $\zeta$ should be detected at 3 standard deviations for variations saturating the current bounds. These forecasts are discussed in \cite[Alves \etal\ (2017)]{Alves}.

In these models the scalar field inevitably couples to nucleons, leading to Weak Equivalence Principle violations. For detailed discussions of this point see \cite[Dvali \& Zaldarriaga (2002)]{Dvali} and \cite[Chiba \& Kohri (2002)]{Chiba}. Hence astrophysical tests of the stability of the fine-structure constant constrain the Eotvos parameter. The current 2-sigma bound for these models is from \cite[Martins \etal\ (2016)]{Pinho},
\begin{equation}
\eta<1.6\times10^{-14}
\end{equation}
which is about an order of magnitude stronger than the ground-based direct bounds, from torsion balance or lunar laser ranging experiments. The MICROSCOPE satellite has recently announced a preliminary bound of $\eta<1.4\times10^{-14}$---at 1 sigma, and expected to improve, see \cite[Touboul \etal\ (2017)]{Microscope}---but high-resolution ultra-stable spectrographs are competitive: the ESPRESSO GTO can reach a sensitivity about 5 times better than MICROSCOPE, while the ELT-HIRES sensitivity is expected to be similar to that of the proposed STEP satellite.

Astrophysical measurements of the fine-structure constant can also be used to reconstruct the dark energy equation of state. Standard methods (such as Type Ia supernovae) are of limited use as dark energy probes: since the field is slow-rolling when dynamically important, a convincing detection of a dynamical equation of state will be difficult at low redshifts. One must probe the deep matter era regime, where the speed of the hypothetical scalar field is likely fastest (even though it is not yet dominating the cosmological dynamics). Next-generation facilities will map the dark side of the universe to $z\sim4$. As discussed in \cite[Leite \etal\ (2014)]{Leite}, constraints comparable to that of a SNAP-like survey can be obtained from fine-structure constant measurements in about 20 nights of ELT-HIRES time. Alternatively, this would take about 700 nights of ESPRESSO time---raising the interesting possibility of building a dedicated post-ESPRESSO spectrograph to use one of the VLT telescopes full-time for a few years.

\section{Synergies and the redshift drift}

In addition to the fundamental nature and direct implications of these measurements, they are also important sources of synergies with other cosmological experiments. In other words, even in the cases where these constraints, on their own are not as stringent as those obtained from other observables, they often probe regions of parameter space that are otherwise inaccessible to other observables, and their combination therefore leads to significantly improved constraints. An example of such a synergy with ESA's Euclid satellite has been discussed in \cite[Calabrese \etal\ (2014)]{Erminia}. The addition of ELT measurements of the fine-structure constant to Euclid constraints on the $w_0$--$w_a$ dark energy parameterisation lead to a small improvement on the $w_0$ constraint, but improves the constraint on $w_a$ by about a factor of 5, due to the much larger redshift lever arm of the former measurements,

\begin{figure}
\begin{center}
 \includegraphics[width=4in]{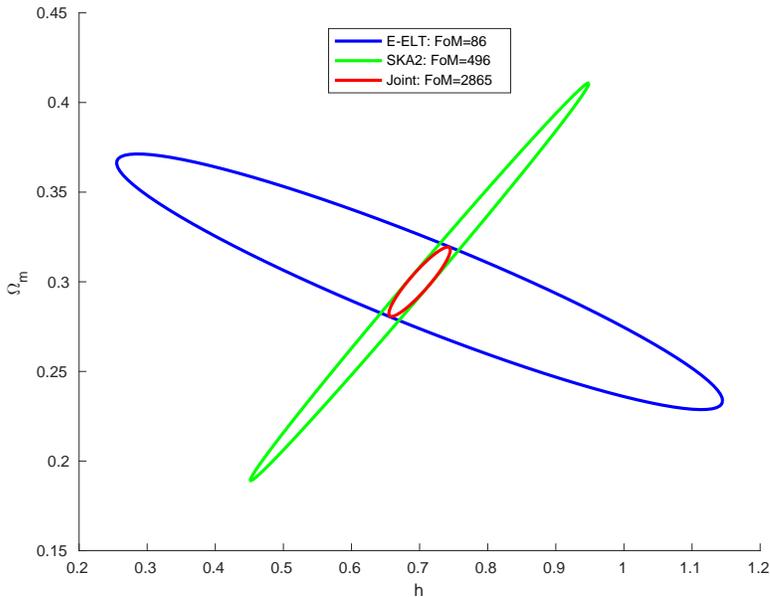}
 \caption{Constraints on the $\Omega_n$--$h$ plane from redshift drift measurements by the ELT and SKA (Phase 2), assuming a flat $\Lambda$CDM model and no external priors. The figures of merit listed in the legend correspond to the inverse of the area of the displayed 1-sigma confidence ellipses.}
   \label{fig1}
\end{center}
\end{figure}

Another example of an astrophysical probe which combines fundamental significance with synergistic value is that of the redshift drift. This is a direct non-geometric model-independent measurement of the universe's expansion history. It is independent of gravity, geometry or clustering: instead of  mapping our (present-day) past light-cone, it directly compares different past light-cones. This is a key ELT-HIRES driver (probing $2<z<5$) as discussed in \cite[Liske \etal\ (2008)]{Liske}, while the full (Phase 2) SKA may measure it at $z<1$, as discussed in \cite[Klockner \etal\ (2015)]{Klockner}. Forecasts for these measurements, using the specifications in the two references, as well as for the combined constraint, are shown in Figure \ref{fig1}, for the pessimistic case where no external priors are used (but assuming the simplest, flat $\Lambda$CDM fiducial model). Further details can be found in \cite[Alves \etal\ (2018a)]{Alves18}.

\section{The blue bottleneck}

An important limiting factor in the above measurements is the access to the blue part of the spectrum (ideally, starting right at the atmospheric cutoff). We will illustrate this with a comparative forecast, assuming as fiducial model a class of so-called Olive-Pospelov models which have been recently studied in \cite[Alves \etal\ (2018b)]{Alves18s}. These models have two free particle physics parameters, which can be constrained using astrophysical measurements of the fine-structure constant. We forecast how well they can be constrained by the ESPRESSO spectrograph, either using only its GTO targets or using all the good targets that ESPRESSO can see (given its wavelength range). This forecast is then repeated for the next-generation high-resolution spectrographs: G-CLEF for the GMT, HROS for the TMT, and ELT-HIRES for the ELT. For the first two we assume the blue cutoffs provided in the publicly available literature, while for ELT-HIRES (whose blue cutoff is not yet decided) we study three possible scenarios. The metric used in the comparison is the {\it figure of merit}, in the usual cosmological constraints sense: the inverse of the area of the 1-sigma confidence ellipse in the plane of the two free parameters. (The overall normalisation of the figure of merit is irrelevant, what is meaningful is the comparison of its value in the different scenarios being studied.)

\begin{figure}[b]
\begin{center}
 \includegraphics[width=4in]{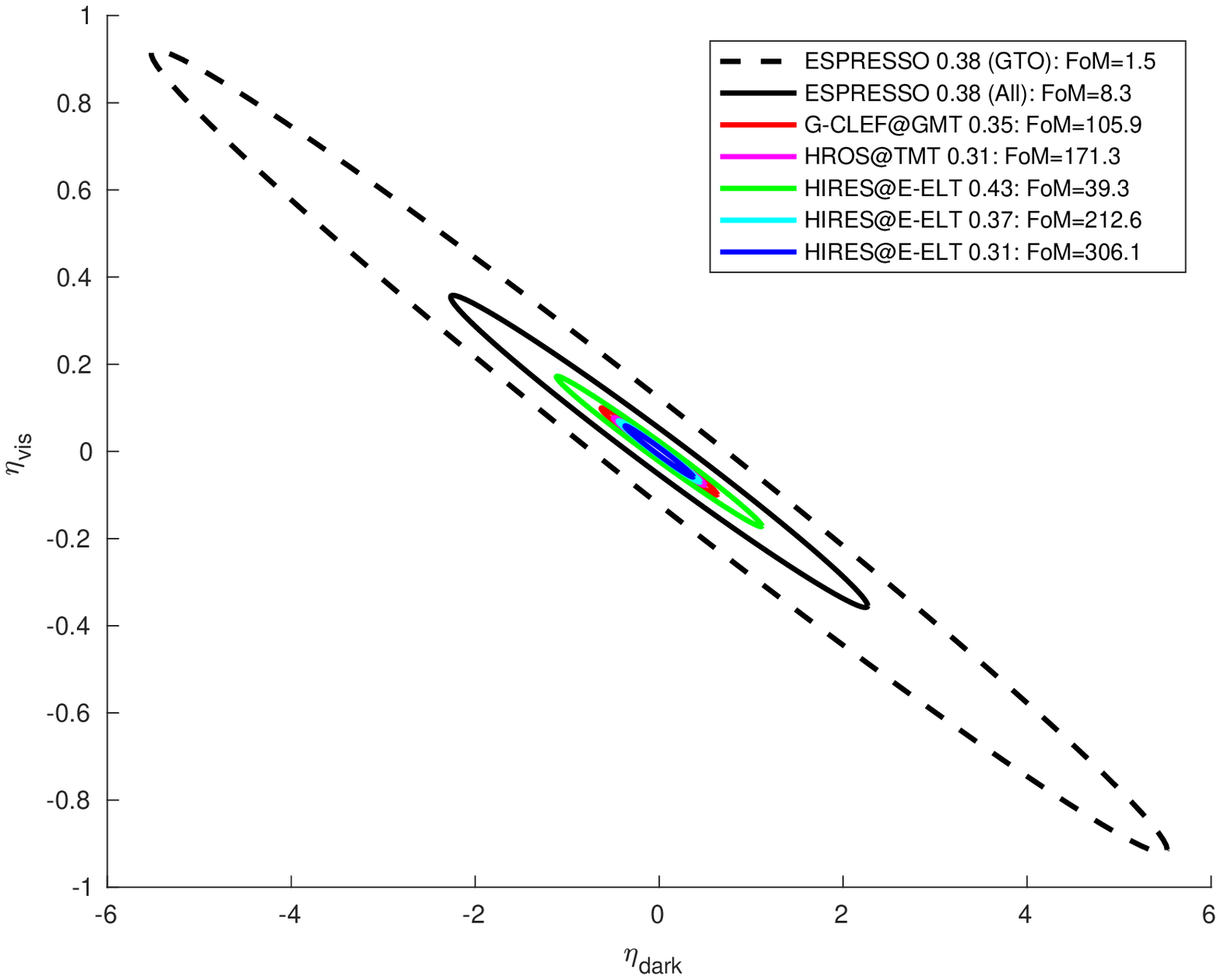}
 \includegraphics[width=4in]{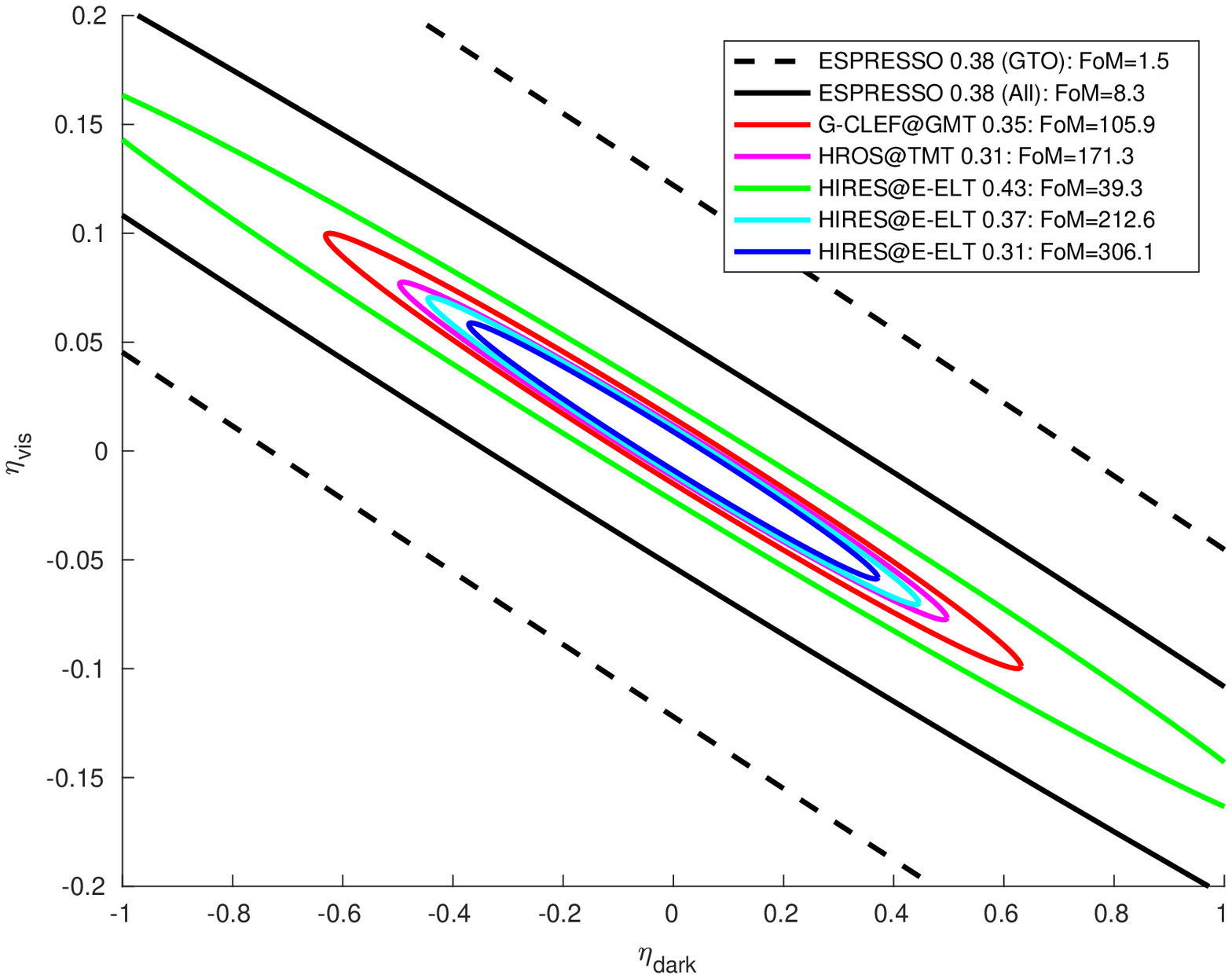}
 \caption{Forecasts of constraints on the two particle physics parameters in the Olive-Pospelov model, for the ESPRESSO spectrographs and the high-resolution spectrographs foreseen for the ELTs, under various assumptions discussed in the text. The figures of merit listed in the legend correspond to the inverse of the area of the displayed 1-sigma confidence ellipses. The bottom panel is a zoomed-in version of the top one.}
   \label{fig2}
\end{center}
\end{figure}

Figure \ref{fig2} shows the results of this analysis. It is clear that for these fundamental physics tests the blue coverage is at least as important as collecting area. The important overall conclusion is that for ELT-HIRES to be competitive with G-CLEF (a first-light instrument) or HROS (a second-generation instrument) it needs to be at least as blue as ESPRESSO---recall that the latter starts at 380 nm.

\section{Conclusions and outlook}

The acceleration of the universe shows that canonical theories of cosmology and particle physics are incomplete, if not incorrect. This contribution highlighted how precision astrophysical spectroscopy provides a direct and competitive probe of the (still unknown) new physics that must be out there, focusing on astrophysical tests of the stability of fundamental dimensionless couplings such as the fine-structure constant and the proton-to-electron mass ratio.

Current data shows that nothing is varying at the few parts per million level. This is already a strong constrain (stronger, for example, than the Cassini spacecraft bound on the behaviour of gravity in the solar system), and it leads to the best available (albeit mildly model-dependent) constraints on Weak Equivalence Principle violations. The arrival of the ESPRESSO spectrograph (whose GTO has just started, at the time of writing) has opened a new era in the field, and significantly improved results will come shortly.

The ELTs have the potential to become the leading gravity and fundamental cosmology probes. Key contributions include Weak Equivalence Principle tests (mostly from measurements of the fine-structure constant), test of composition-dependent force tests (mostly from measurements of the proton-to-electron mass ratio), mapping the dark side of the Universe from $z=0$ to $z=4$, and a direct model-independent probe of the universe dynamics (through the redshift drift measurements, which are within the reach of the ELT, though likely not the TMT or the GMT). Further possibilities not discussed in this contributions include improved measurements of the primordial Deuterium abundance, strong-field tests of gravity (including the so-called No-hair Theorem), and probing the weak acceleration MOND-like regime in the outskirts of the Milky Way.

The requirements include 50 to 250 nights of telescope time over the instruments' lifetime (depending on how much of the above portfolio one subscribes to), the identification of further 'clean' targets (especially for measurements of the proton-to-electron mass ratio and the redshift drift), and improved measurements of the laboratory (rest) wavelengths of most of the relevant atomic and molecular transitions. (In the above I have assumed that one is only interested in redshifts up to $z=4$; in principle one can go beyond this, either by going into the infra-red or by using transitions below 160 nm---although these are not well known in the lab.) Last but by no means least, reasonable access to UV/blue wavelengths is essential.

The ELTs will therefore be the flagship tool in a new generation of precision consistency tests of fundamental cosmology, leading to competitive guaranteed science implications for dark energy and fundamental physics: any new and improved measurement will rule out some previously allowed classes of theories or regions of parameter space, even if they are null results. They also possess a unique value of complementarity, redundancy, and synergies with other facilities including ALMA, Euclid and SKA (and many of these synergies remain unexplored).

This work was financed by FEDER---Fundo Europeu de Desenvolvimento
Regional funds through the COMPETE 2020---Operational Programme for
Competitiveness and Internationalisation (POCI), and by Portuguese
funds through FCT---Funda\c c\~ao para a Ci\^encia e a Tecnologia in the
framework of the project POCI-01-0145-FEDER-028987.

Many interesting discussions with other members of the CAUP Dark Side team (Ana
Catarina Leite, Ana Marta Pinho, Catarina Alves, Jo´s\'e Guilherme Matos,
Maria Carolina Faria and Tom\'as Silva) as well as with many other colleagues and
collaborators in the work discussed herein (Erminia Calabrese, Gemma Luzzi, Hugo
Messias, Joe Liske, John Webb, Matteo Martinelli, Paolo Molaro, Ricardo
G\'enova-Santos, Stefano Cristiani and Tasos Avgoustidis) are gratefully acknowledged.

\begin{discussion}

\discuss{Cai}{On the forecasted constraints for the redshift drift in the $\Omega_m$--$h$ plane, why are the directions of degeneracy completely different for the ELT and the SKA2?}

\discuss{Martins}{The reason is that the ELT can measure the redshift drift at redshift $z>2$ (in the deep matter era) while the SKA can only do it at $z<1$ (in or very near the acceleration phase). The fact that they are probing the dynamics of the universe at such different redshifts allows the combination of the two to break parameter degeneracies.}

\end{discussion}

\end{document}